\documentclass[aps,prl,11pt]{revtex4}
\usepackage{amssymb,epsf}

\usepackage{latexsym}
\usepackage{xcolor}
\usepackage{epsfig}
\usepackage{float} 
\usepackage{color} 
\usepackage{amsmath}
\usepackage{bm}
\usepackage{amsthm}
\usepackage{amssymb}
\usepackage{amssymb,epsf}
\usepackage{latexsym}
\usepackage{epsfig}
\usepackage{graphicx}

\begin{document}

\title{Effective information bounds in modified quantum mechanics} 

\author{Sarah Aghababaei$^1$\footnote{sarah.aghababaei@gmail.com}, Hooman Moradpour$^2$\footnote{hn.moradpour@maragheh.ac.ir},  Salman Sajad Wani$^3$\footnote{sawa54992@hbku.edu.qa}, Francesco Marino$^4$ \footnote{francesco.marino@ino.cnr.it}, Naveed Ahmad Shah$^{5,6}$\footnote{naveed179755@st.jmi.ac.in},  Mir Faizal$^{6}$\footnote{mirfaizalmir@gmail.com}}

\affiliation{$^1$Department of Science, University of Farhangian, Yasouj Branch, Iran\\
    $^2$Research Institute for Astronomy and Astrophysics of Maragha (RIAAM), University of Maragheh, 55136-553, Maragheh, Iran\\
    $^3$Qatar Center for Quantum Computing, Hamad Bin Khalifa University, Doha, Qatar\\
      $^4$CNR-Istituto Nazionale di Ottica and INFN Sezione di Firenze, Via Sansone 1, I-50019 Sesto Fiorentino (FI), Italy.\\
    $^5$Department of Physics, Jamia Millia Islamia, New Delhi - 110025, India\\
    $^6$Canadian Quantum Research Center (CQRC)
    204-3002 32 Ave Vernon, BC V1T 2L7 Canada}

\begin{abstract}

A common feature of collapse models and an expected signature of the quantization of gravity at energies well below the Planck scale is the deviation from ordinary quantum-mechanical behavior. Here, we analyze the general consequences of such modifications from the point of view of quantum information theory and we anticipate applications to different quantum systems. We show that quantum systems undergo corrections to the quantum speed limit which, in turn, imply the modification of the Heisenberg limit for parameter estimation. Our results hold for a wide class of scenarios beyond ordinary quantum mechanics. For some nonlocal models inspired by quantum gravity, the bounds are found to oscillate in time, an effect that could be tested in future high-precision quantum experiments.

\end{abstract}
\maketitle

\section{Introduction}

Quantum mechanics has proven to be extraordinary powerful to provide an accurate description of physical processes at the atomic and subatomic scale and has enabled the development of new technologies that are now part of daily life. In spite of roughly a century of predictive success, there are still hot debates about its conceptual backgrounds, in particular concerning the consistency and completeness of the theory. A notable example is the measurement problem, namely, how a superposition of states collapses towards a definite outcome. The concept of measurement-induced collapse of the wave function, known as the Copenhagen interpretation, provides the classical solution to this issue. An alternative approach is given by models of wave-function collapse (collapse models) that instead consider modifications of the theory. In such models the usual linear and deterministic Schr\"{o}dinger evolution is modified by the addition of nonlinear and stochastic terms, in an attempt to include in the theory the intrinsic nonlinearity and stochasticity of the measurement process \cite{ghirardi86,ghirardi90,bassi2003,bassi2013}. Other more specific models have been proposed in literature, the most notable being the one by Diosi and Penrose for the gravitational-induced collapse \cite{penrose96,diosi}. 
While these modifications produce the spontaneous collapse of the wave function independently of any measurement processes, they also lead to new phenomena in appropriate experimental regimes and several precision experiments have been recently performed in search of these effects \cite{1a,1b,vinante19,curceanu}. 

Alternative models for gravitational decoherence rely on the deformation of the Heisenberg algebra and the corresponding modification of the uncertainty principle \cite{class}. Such generalized uncertainty relations arise from the prediction, common to string theory, loop quantum gravity and Gedanken experiments in black hole physics, of the existence of a minimal observable length \cite{veneziano,gross,sabine,garay,scardigli}. This has suggested the possibility to detect low-energy signatures of quantum gravitational effects in the form of deviations from the standard quantum theory \cite{ml12}. Proposed settings include accurate spectroscopic measurements \cite{ml15,space,space1}, gravitational bar detectors \cite{ab12} and optomechanical experiments \cite{ml14,omt}, to name just a few. 

Remarkably, modifications of quantum mechanics by a deformation of the Heisenberg algebra are not limited to realm of quantum gravity, but naturally arise as higher-order corrections in any low-energy effective field theory \cite{effect1,effect2}. Such modified theories provide a description of the dynamics of quantum fluctuations in several condensed-matter systems \cite{girelli08,marino19,gupa,gupb} that could be thus exploited for simulating these effects in the laboratory.

All the aforementioned deviations generally imply changes in the energy spectrum of quantum systems, as well as in the time evolution, expectation values, and uncertainties of a given observable. These effects of these modifications have been analysed in different contexts in the last decades (see e.g. Refs. \cite{ml12,ml15,space,kempf,chang,lewis,ching,pedram,non1,non2}), in particular to shed light on detectable infrared effects of quantum gravity, or to put constraints on the related deviation parameters. Surprisingly, however, the general implications of modified quantum rules on quantum metrology and quantum computation have not been investigated so far. 

In ordinary quantum mechanics, the minimum time required for a quantum system to evolve between two distinguishable states is limited by the quantum speed limit \cite{qsl}. This limit is closely related to time-energy uncertainty relations as illustrated by the Mandelstam-Tamm bound \cite{mt}. The quantum speed limit can be also expressed in terms of the average energy of the system via the Margolus-Levitin bound \cite{ml}. The latter can also be identified with the Heisenberg limit for parameter estimation and, as demonstrated by Lloyd, can be used to derive a fundamental upper limit in the computation speed \cite{compution}. Remarkably, it was shown that environmental effects, such as non-Markovian dynamics can speed up quantum processes \cite{deffner}, which was also verified in a cavity quantum electrodynamic experiment \cite{cimmarusti}. This motivates the question about the universality of such fundamental limits and whether possible violations could be the manifestation of some new physics coming into play.

In this paper, we show that an increase in the quantum speed limit is expected for a wide class of deviations from quantum mechanics, as those predicted by theories that aim to merge quantum physics and gravity. Since in all approaches
these deviations are proportional to the ratio between the energy scale probed by the system and the scale at which conventional quantum theory ceases to be applicable, they manifest as small corrections to the standard quantum dynamics, which justify the use of a perturbative expansion of usual quantum operators and states. We derive generalized expressions of the above fundamental bounds and discuss how these imply the lowering of the Heisenberg limit and the increase of Lloyd's bound on the maximum speed at which quantum computation can be performed. 
We suggest that since black holes attain the Lloyd's computation bound, analogue black holes in condensed-matter experiments could be used to realize, at least in principle, quantum computation at a speed saturating the modified bounds emerging in their systems. Our results are general as they do not rely on any specific modification of quantum mechanics. We finally focus on a specific class of quantum gravity-induced non-local modifications of quantum mechanics and predict a scenario in which the bounds are oscillating in time, an effect that could be testable in future high-precision experiments on macroscopic quantum oscillators. 

\section{Theoretical Framework}

We start our analysis from an operator $K$, which is the generator of translations in the parameter $\theta$ and denote its eigenstates by $ | i \rangle$. When $K$ is for instance the Hamiltonian operator $H$, the parameter $\theta$ corresponds to the time variable $t$. 
We remark that this formalism can be applied to any general operator probing a parameter like time or phase, but cannot be directly applied to study the relation between two operators, like e.g. position and momentum. We consider a generic modification of the quantum theory such that the eigenstates of the operator $K$ will be modified to $ | i_g \rangle$. For consistency the deviations are assumed to be small in which case the generalized eigenstate can be expressed in terms of the original eigenstates $ | i \rangle$ and a first-order correction $| i_1 \rangle$ as $| i_g \rangle =   | i \rangle + \eta | i_1 \rangle$, with $\eta$ being a dimensionless parameter quantifying the deviation for the ordinary quantum behaviour. Similarly, we write the new state of the system by $| \Psi_g \rangle $, such that $| \Psi_g \rangle = | \Psi \rangle + \eta | \Psi_1 \rangle$.

The sign of the parameter $\eta$ depends on the specific model under consideration. However, for most for most of the above theories, which achieve a modification to standard quantum mechanics due to gravitational effects (see e.g. \cite{ml12,bassi2013,kempf} ), $\eta$ is positive, and a further example is provided by the nonlocal scenario analyzed in this work. In this manuscript, we will primarily concentrate on this class of theories, bearing in mind that negative values of $\eta$ are also feasible.

It is important to highlight that the above expansions should not be interpreted as usual perturbed states (for instance, as those associated to some complex Hamiltonian) in ordinary quantum mechanics, but rather as an effective first-order description of the dynamics of a quantum system that obeys some modified quantum rules, where $\eta$ encodes the scale at which such modifications come into play. We will revisit this concept later on.

Within this generalized framework, and at the first order in $\eta$, the probability for the state $ | \Psi_g \rangle $ to collapse into a state $ | i_g \rangle $ 
can be as $  P_{ig}$ can be expressed in terms of the original probability $P_i$ for the state $| \Psi \rangle $ to collapse into $| i \rangle$ and its first order corrections $P_{i1} $, such that 
$ P_{ig}=   P_{i} + \eta P_{i1} $. 
We can express the generalized density matrix  $\rho_g =\sum_{i}p_{ig}|i_g\rangle\langle i_g| $ as 
$
    {\rho}_g =\sum_{i}p_i|i\rangle\langle i|+\eta(p_i|i\rangle\langle i_1|+ p_i|i_1\rangle\langle i|+p_{1i}|i\rangle\langle i|)= {\rho}+\eta  {\rho}_1 
$ and define  $\rho'_g  = d\rho_g/d\theta$. Since the generalized quantum Fisher information for the above state is
$  F_{g}(\theta) = Tr( \rho'_g \mathcal{L}_{p_g} (\rho'_g)), $ where  $ 
 \mathcal{L}_{p_g} (\rho'_g) = 2 \sum_{j,k} {(\rho'_g)_{j,k}}({p_{g,j} + p_{g,k}})^{-1} |j_g \rangle \langle k_g| 
$,  we obtain 
\begin{equation}
    F_{g}(\theta) = \frac{2}{\hbar^2} \sum_{j,k} \frac{(p_{g,j} - p_{g,k})^2}{p_{g,j} + p_{g,k}} |\Delta_g (K_g)_{jk}|^2
    \label{eq1}
\end{equation}
The modified variation $ \Delta_g K_g$ can be obtained by deforming both $K$ and the $\rho$. So, it can be expressed in terms of the original variation $\Delta K$ and its first order corrections $\Delta_1 K_1 $ as $ \Delta_g K_g  = \Delta K + \eta \Delta_1 K_1$ because  $\Delta_g K_g = K_g - Tr(\rho_g K_g) =   K + \eta K_1 - Tr(\rho K) - \eta (Tr (\rho K_1) +Tr (\rho_1 K))$. 
Here,  $ \Delta_1 K_1 $ can be expressed as $ \Delta_1 K_1 = K_1 - Tr(\rho K_1) - Tr(\rho_1 K) =  K_1 - \langle K_1 \rangle - \langle K \rangle_1 
 $. Substituting these expressions in Eq. (\ref{eq1}), the generalized quantum Fisher information for the parameter $\theta$ associated with this generator can be written as 
  \begin{eqnarray}
    F_{g}(\theta) &=& \frac{2}{\hbar^2} \sum_{j,k} \frac{((p_{j} - p_{k})^2+ 2\eta( p_j p_{1,j} + p_k p_{1,k}-p_{1,j}p_k-p_j p_{1,k} ))}{(p_{j} + p_{k}) + \eta (p_{1,j} + p_{1,k}) } |\Delta K_{jk} + \eta \Delta_1 K_{1,jk}|^2 \nonumber \\
    &=& \frac{2}{\hbar^2} \left(\sum_{j,k}\frac{(p_{j} - p_{k})^2}{p_{j} + p_{k}} |\Delta K_{jk}|^2 + \eta \sum_{j,k} \frac{(p_{j} - p_{k})^2}{p_{j} + p_{k}} |\Delta K_{jk} \Delta_1 K_{1,jk} | \right. \\
 && \left. + 2\eta \sum_{j,k} \frac{2(p_j +p_k)(p_j p_{1,j}+ p_{1,k}p_k-p_{1,j}p_k-p_j p_{1,k}) - (p_j -p_k)^2 (p_{1,j} + p_{1,k})}{(p_j + p_k)^2}
|\Delta K_{jk}|^2 \right) \nonumber
\end{eqnarray} 
The generalized quantum Fisher information for the parameter $\theta$ can be written in terms of the average of the modified variation $\Delta_g K_g$. Here we have to take the average with respect to the modified eigenstates of $K_g$ as $\langle (\Delta_g K_g)\rangle_g =  Tr ( ({\rho} + \eta {\rho_1}) (\Delta_g K_g))$. As the average is taken over the modified eigenstates, the trace is taken with respect to $({\rho} + \eta {\rho_1}) $ rather than ${\rho}$.  Thus, we can write 
 $F_{g}(\theta) = {4} \langle (\Delta_g K_g)\rangle_g/ {\hbar^2}$. 
Here, using $  \Delta_g K_g =  \Delta K + \eta \Delta_1 K_1  $, up to the first order in $\eta$, we can write 
 \begin{eqnarray}
  \langle (\Delta_g K_g)^2 \rangle_g &=&  Tr \Big( ( {\rho} + \eta  {\rho_1}) ( \Delta K^2 + \eta (\Delta K \Delta_1 K_1 + \Delta_1 K_1 \Delta K))\Big)\nonumber \\
  &= & Tr \ ( {\rho} (\Delta K)^2) + \eta \ Tr  (\rho (\Delta_1 K_1 \Delta K + \Delta K \Delta_1 K_1))  + \eta  \ Tr (\rho_1 (\Delta K)^2)\nonumber \\
  &=& \langle(\Delta K)^2 \rangle+ \eta \langle\Delta K_1 \Delta K + \Delta K \Delta_1 K_1 \rangle+ \eta \langle(\Delta K)^2 \rangle_1
  \end{eqnarray}
  where $\langle(\Delta K)^2 \rangle_1$ denotes the average is taken with respect to $\rho_1$. 
  
For any couple of quantum states, $|\Psi\rangle$  and $|\Phi\rangle$ in a Hilbert space, the statistical distance between them can be expressed in terms of the angle $s=\arccos |\langle \Psi |\Phi \rangle|$ that can be related to the quantum Fisher information through the expression $ F(\theta) = ({ds}/{d\theta})^2$. A similar relation holds between the generalized quantum Fisher information $F_g(\theta)$ and the generalized statistical distance $s_g=\arccos |\langle \Psi_g |\Phi_g \rangle|$, $F_g(\theta) = ({ds_g}/{d\theta})^2$, where $|\Psi_g\rangle$ and $|\Phi_g\rangle$ are the modified states of the system. Using this relation, we can express the bound on the generalized quantum Fisher information as 
  \begin{equation}
      F_g(\theta)   \leq \frac{4}{\hbar^2} \langle (\Delta_g K_g)^2 \rangle_g
  \end{equation}
Integrating this expression, we obtain a generalized  bound on $\theta$ 
  \begin{equation}
      \theta_{min}  \geq \frac{\pi \hbar}{ 2} \left( \frac{1}{\sqrt{\langle (\Delta K)^2 \rangle} }- \frac{\eta (\langle \Delta K_1 \Delta K \rangle + \langle (\Delta K )^2\rangle_1 + \langle \Delta K \Delta K_1\rangle)}{2 \langle (\Delta K)^2 \rangle^{3/2} }\right)
      \label{mb}
  \end{equation}
For the special case in which and $K$ is the Hamiltonian operator $H$ and thus $\theta$ is the time $t$, Eq. (\ref{mb}) gives the modified Mandelstam-Tamm bound 
  \begin{equation}
      t_{min} \geq \frac{\pi \hbar}{ 2} \left( \frac{1}{\sqrt{\langle (\Delta H)^2 \rangle} }- \frac{\eta (\langle \Delta H_1 \Delta H \rangle + \langle (\Delta H )^2\rangle_1 + \langle \Delta H \Delta H_1\rangle)}{2 \langle (\Delta H)^2 \rangle^{3/2} }\right)
  \label{mmt}
  \end{equation}
  
We remark that the above expansions should be interpreted as an effective description of the consequences of some modified quantum rules. Within this generalized framework, and at the first order in $\eta$, a system governed by the Hamiltonian $H$ evolves effectively as if it were governed by the Hamiltonian $H + \eta H_1$ in ordinary quantum mechanics. Consequently, it undergoes a modified limit \ref{mmt}, which mirrors the limit of a system governed by the Hamiltonian $H$ in standard quantum mechanics.
  
One can readily verify that Eq. (\ref{mmt}) reduces to the standard bound \cite{mt} in the limit $\eta \to 0$, which can be readily put in the form of a time-energy uncertainty relation. The modified bound (\ref{mmt}) can be thus also viewed as the consequence of a generalized uncertainty relation emerging in the context of a modified Schr\"{o}dinger evolution. On the other hand, the above expression is not a statement about simultaneous measurements, but it should be rather interpreted as the time a quantum system needs to evolve from an initial to a final state.

This interpretation was further supported by Margolus and Levitin's result on the minimal time for a quantum system to evolve to an orthogonal state. To calculate this quantity in our generalized framework we first observe that starting from  $|\Psi_{g,0}\rangle  $ at $\theta =0$ we can evolve it to $|\Psi_{g, \theta}\rangle$ as 
  \begin{equation}
  |\Psi_{g,\theta}\rangle=e^{-{i} K_g\theta/ \hbar}|\Psi_{g,0}\rangle \,
  \label{ev}
 \end{equation}
where $K_g$ is a generic operator. As before, the generalized statistical distance between the quantum states at $\theta=0$ and any finite $\theta$ is $s_{g} =\arccos |\langle \Psi_{g,0} |\Psi_{g,\theta} \rangle|$.  Using (\ref{ev}), we observe that $- d   |\langle \Psi_{g,0} |\Psi_{g,\theta}  \ \rangle| /d\theta \leq  | d  \langle \Psi_{g,0} |\Psi_{g,\theta}  \rangle /d\theta  |$ and from Schwarz inequality we obtain \cite{Jones2010} 
\begin{equation}
\frac{d s_g}{d\theta}\leq\Big|\frac{d}{d\theta}|\langle\Psi_{g,0}|K_g|\Psi_{{g,\theta}}\rangle|\Big|
\leq \frac{|\langle\Psi_{g,0}|K_g|\Psi_{{g,\theta}}\rangle|}{\hbar}
\leq \frac{|\langle\Psi_{g,0}|K_g |\Psi_{{g,0}}\rangle|}{\hbar}
\equiv \frac{|\langle K_g\rangle_g|}{\hbar}
\end{equation}
Integrating this equation, we get the deformed bound for $\theta$ in terms of average of the operator $\theta\geq {\hbar}{\pi}/2{|\langle K_g\rangle_g|}
={\hbar\pi}/2{|Tr( {\rho}_gK_g)|}$. Expanding up to the first order in $\eta$ yields 
  \begin{equation}
     \theta_{min} \geq \frac{\hbar\pi}{2}\Bigg|\frac{1}{\langle K \rangle}-\eta\bigg(\frac{\langle K_1\rangle+\langle K\rangle_1}{\langle K\rangle^2}\Bigg)\Bigg|
     \label{mml}
 \end{equation} 
When $K$ is the Hamiltonian operator $H$ and $\theta$ the time variable $t$, Eq. (\ref{mml}) takes the form of a generalized Margolus-Levitin bound  
\begin{equation}
   t_{min} \geq \frac{\hbar\pi}{2}\Bigg|\frac{1}{\langle H \rangle}-\eta\bigg(\frac{\langle H_1\rangle+\langle H\rangle_1}{\langle H\rangle^2}\Bigg)\Bigg| 
   \label{mlbt}
 \end{equation} 
which reduces to the standard result when $\eta \to 0$ \cite{ml}.  

\section{Physical implications: from quantum metrology to quantum computing}

The above modification can have important physical consequences, for instance in the dynamics of the orthogonality catastrophe \cite{orth12}, the classification of symmetry-breaking properties of states and operations \cite{symme}, the Kibble-Zurek mechanism \cite{kz12}, to name just a few. 
Maybe one of the most intriguing features is that quantum speed limit is related to spacetime quantum fluctuations  \cite{2001a}. An increased quantum speed limit is expected to imply the amplification of such fluctuations, with implications ranging from the cosmological constant \cite{foam1, foam5, foam2, foam4} to the causal structure of spacetime \cite{1995}. Remarkably, such amplified fluctuations should translate into an additional source of noise in highly-sensitive interferometric experiments that, in principle, could be detected or constrained by gravitational wave detectors \cite{ligo1, ligo2}.

A further relevant question related to the above modified fluctuations is whether they increase or decrease the accuracy of measurement of a parameter probed by a modified operator, such as time \cite{miro1, miro9} or phase \cite{miro2, miro4, miro6}. This issue is addressed by investigation the modifications to the Heisenberg limit, which ultimately bounds the accuracy of measurement of such parameters \cite{quant1}.

The Margolus-Levitin bound can be in fact identified with the Heisenberg limit, which has been shown to be optimal for all parameter estimation procedures in quantum metrology \cite{quant1}. 
Using the modified bound (\ref{mml}), we can write the generalized Heisenberg limit as 
  \begin{equation}
   \delta  \theta \geq \hbar\Bigg|\frac{1}{\langle K \rangle}-\eta\bigg(\frac{\langle K_1\rangle+\langle K\rangle_1}{\langle K\rangle^2}\Bigg)\Bigg| \ ,
     \label{mml2}
 \end{equation} 
which for $\eta>0$ corresponds to an increase of the achievable precision in the estimation of $\theta$.

A further fundamental bound which highlights the deep relation between metrology and quantum speed limits is the Cram\'{e}r-Rao lower bound for the variance of an unbiased estimator \cite{est1, est2}. Defining $\delta \theta$ as the root mean square error in the estimation of $\theta$, we have $\delta \theta \geq 1/ \sqrt{\nu F(\theta)}$ \cite{nu}, where $\nu$ the number of times the system has been probed and $F(\theta)$ the quantum Fisher information. Any generalization of the quantum Fisher information, $F_g(\theta)$ will thus modify the Cram\'{e}r-Rao bound to 
\begin{equation}
\delta \theta \geq 1/ \sqrt{\nu F_g(\theta)} \ .
\label{mcr}
\end{equation}

If we consider that the parameter we wish to estimate is the elapsed time from a given evolution governed by a time-independent Hamiltonian $H$, then Eq. (\ref{mcr}) bounds the time uncertainty by the modified quantum Fisher information. 

In computing language, the Margolus-Levitin's theorem on the maximum rate at which a quantum system can evolve through distinct states, translates into a theoretical upper limit on the number of operations that can be performed in a unit time. Lloyd applied Margolus–Levitin's results to derive an upper bound for the computation speed of any system and argued that the maximum rate is actually attained by black holes \cite{compution}. The saturation of the Lloyd's bound by black holes has been then rigorously demonstrated using quantum complexity and its relation to fidelity \cite{qc1, qc2}. If black holes can contain hairs, e.g. quantum fields in equilibrium with the black hole, neither infalling nor escaping to infinity, such configurations would conceptually represent the fastest achievable quantum computers. In the presence of modifications of quantum mechanics (e.g. induced by quantum gravitational effects), the corresponding bounds would be even lower due to the increased quantum speed limit.

While a real black hole is certainly not the most appropriate platform one could think to perform computations or experimental tests, the above ideas have the merit to provide new insights on the properties that a hypothetical system should have in order to attain these limits. The so-called analogue-gravity models, in which Lorentz invariance is recovered as a low-energy limit of modified dispersion relations has been successfully used in the last decade to simulate black hole geometries and related phenomena, such as the emission of Hawking radiation and Penrose superradiance \cite{rev}. The scale at which the higher-order corrections become effective and the Lorentz symmetry is broken is related to the underlying microscopic physics of the specific system. Numerical and experimental studies based on these models have demonstrated the robustness of the Hawking emission \cite{carusotto2008,steinhauer2014,steinhauer2016,weinfurtner2011,drori} and the Penrose process \cite{prain2019,torres,braidotti2022} against the short-scale cutoff of field modes, thus supporting their interpretation as infrared effects. 

On the other hand, these systems represent a natural platform to test fields on black-hole spacetime obeying modified quantum laws. At difference with quantum gravitational ultraviolet corrections to real black holes, the way in which such corrections emerge from the microscopic theory in analogue-models is reasonably well modelled and understood. Remarkably, in some of these models the elementary collective excitations evolve as massive scalar fields that in the non-relativistic limit are governed by modified Schr\"{o}dinger equations \cite{girelli08,marino19} similar to those of collapse models. In addition, analogue of hairy black hole solutions arise in these systems as equilibrium configurations between the massive excitation field and the black-hole geometry \cite{ciszak,hod}. Although a precise correspondence of these solutions with those found in General Relativity or modified gravity theories cannot be established, it would be anyway fascinating to explore these systems to test novel quantum computation limits in black hole scenarios beyond ordinary quantum mechanics.

\section{A specific example: non-local modifications of quantum mechanics}

We now analyze the implications of our results on a specific class of non-local modifications of quantum mechanics \cite{belen16,belen17}. These modifications are consistent with the low-energy predictions of non-local effective field theories arising in different approaches to quantum gravity. Examples are found in the context of string field theory \cite{eliezer}, non-commutative geometries \cite{szabo}, causal set theory \cite{sorkin} and loop quantum gravity \cite{gambini}. Such non-local effective field theories lead to a modified Schr\"{o}dinger evolution in the nonrelativistic limit \cite{belen16,belen17}. High-precision experiments in quantum systems could thus offer a concrete possibility of observing quantum gravity effects, or at least constrain a whole class of candidate theories. 
   
Most of on-going experiments and recent proposals are based on optomechanical systems, where deviations from the standard behaviour could be detected in the dynamics of macroscopic mechanical oscillators \cite{ml14,omt,belen17}. Motivated by these investigations, we focus on the non-local quantum evolution of a harmonic oscillator. The non-local Hamiltonian, truncated at the first order in the dimensionless parameter $\eta$ is given by \cite{belen19}
\begin{equation}
 H_g =   H+ \eta H_1= \frac{1}{2}\left({x}^2+ {p}^2\right) + \eta \left( \frac{{x}^4}{2} + 2 i x p + 1 \right) 
 \label{ham}
\end{equation}
where here $\sqrt(\eta)$ is proportional to the ratio between the scale of non-locality and the width of the oscillators
ground-state wave function (see also Appendix for more details). The average energy calculated on a generic state can be expressed as  
  \begin{eqnarray} 
  \langle H_g \rangle_g &=& \frac{1}{N} \left(
    \langle\psi(x,t)| H|\psi(x,t)\rangle \right) \nonumber \\
    &+& \frac{\eta}{N}\left(Re[\langle\psi(x,t)|  H_1 |\psi(x,t)\rangle] + \langle \psi_1(x,t)| H|\psi(x,t)\rangle+ \langle\psi(x,t)| H|\psi_1(x,t)\rangle\right) 
    \label{hav}
\end{eqnarray}
where we used the ansatz $| \Psi_g \rangle =  | \Psi \rangle + \eta | \Psi_1 \rangle$ and we neglected second-order terms in $\eta$.

\begin{figure}
    \centering
\includegraphics[width=0.5\textwidth]{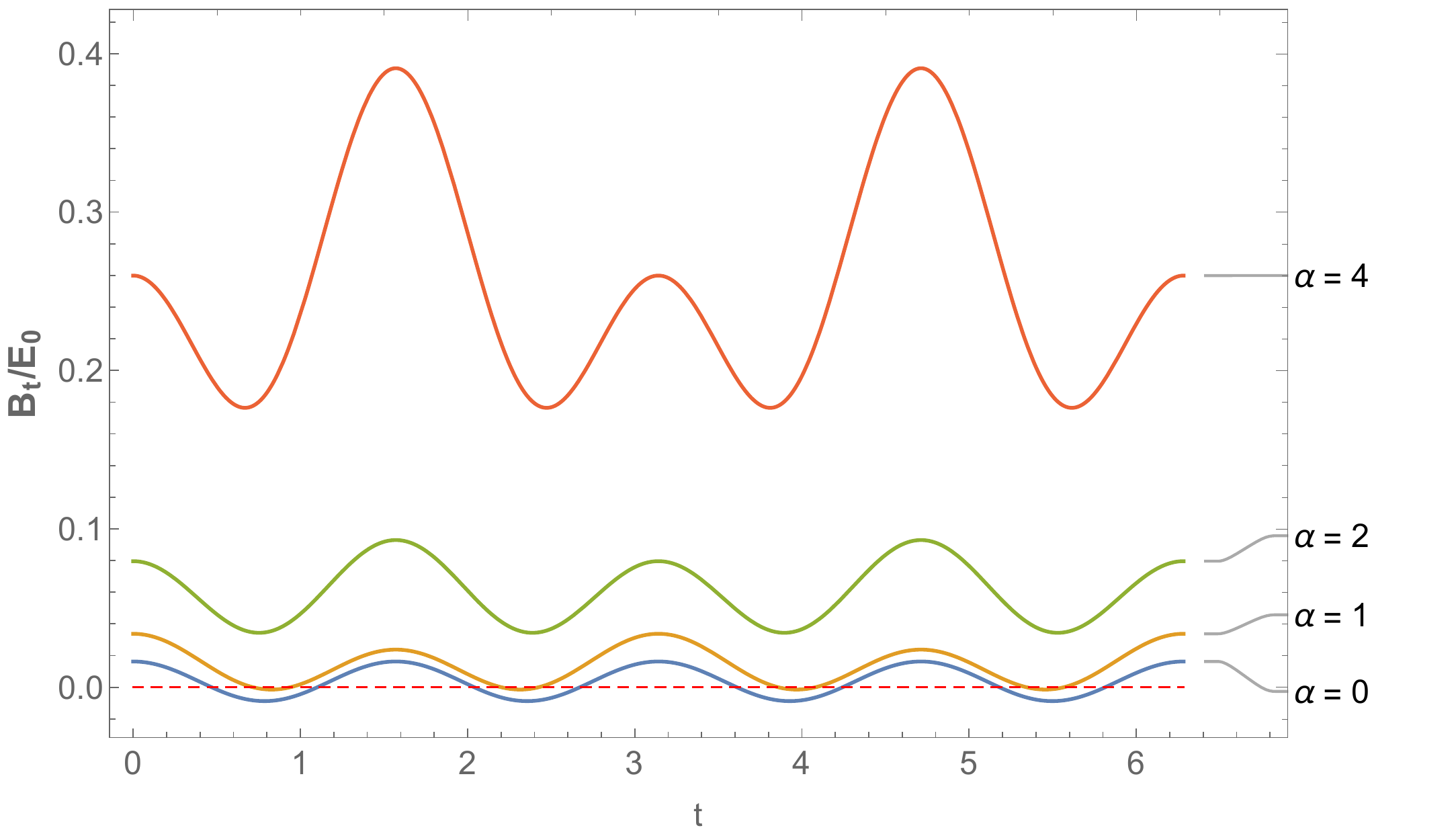}\caption{The correction term $B_t/E_0$ to the limits (\ref{bound}) and (\ref{bound2}) as a function of time for different values of the coherent amplitude and $\eta=0.01$.}% All energy components in the plot are normalized to the unperturbed value $E_0$ (dashed horizontal line).}
    \label{fig1}
\end{figure}

We consider $|\psi(x,t)\rangle$ to be a generic coherent state of amplitude $\alpha$ (the corresponding results for the ground state can be found by setting the coherent amplitude $\alpha = 0$). This choice is motivated by the fact that the preparation of a mechanical oscillator in a pure state (e.g a Fock state) is not straightforward and has never been achieved so far in optomechanics. On the other hand, mechanical oscillators have been already cooled down to thermal occupation numbers below unity, i.e. close to the quantum ground state (see e.g. \cite{peterson,Qiu}) and can be excited in a state of definite coherent amplitude. For such a state Eq. (\ref{hav}) can be written as $\langle H_g \rangle_g \equiv B(t) \approx E_0 + \eta B_t$ where $B_t$ is a periodic function of time (see Appendix). Using (\ref{mml}) we can derive a minimum time $t_{min}$ for a coherent state to evolve into another given by \cite{note}
\begin{equation} 
t_{min} \geq \frac{\pi \hbar }{2 B(t)} \approx \frac{\pi \hbar }{2 E_0} \left(1-\eta \frac{B_t}{E_0} \right) \, .
\label{bound}
\end{equation}
The correction $B_t/E_0$ for some values of the coherent amplitude $\alpha$ is plotted in Fig. \ref{fig1}.
For higher values of $\alpha$, the curves are everywhere positive, indicating that the minimum time $t_{min}$ always falls below the standard Margolus-Levitin limit (Eq. (\ref{bound})). Conversely, at lower $\alpha$ values, in particular at the ground state ($\alpha=0$), the $B_t/E_0$ ratio periodically drops below zero. As a result, the minimum time periodically exceeds the standard limit, albeit exhibiting a lower average value. While the latter result was indeed predicted by the general formula (\ref{mml}) for $\eta >0$, the time dependence of the bound is a specific consequence of the model here considered, and can be attributed to first-order effects of non-locality which are mainly encoded in the anti-Hermitian part of the Hamiltonian (\ref{ham}), $2 i x p + 1$. Using Eq. (\ref{mml2}), with $\theta=t$ and $\langle K_g \rangle_g = \langle H_g \rangle_g$, we obtain \begin{equation}
   \delta t \geq \frac{ \hbar}{E_0}\left(1-\eta\frac{B_t}{E_0}\right) \ . 
\label{bound2}
\end{equation}
which provide a correction to the Heisenberg limit. Therefore, even the accuracy in measuring time $\delta t$ now becomes a function of the time itself, a result that is specific of this form of non-locality at least at the level of first-order perturbative terms.

\section{Conclusions}

In conclusion, we demonstrated that for a wide class of modifications of quantum mechanics as, for instance, those expected as consequences of quantum gravitational effects, the quantum speed limit is increased. We derived generalized expressions for the Mandelstam-Tamm, the Margolus-Levitin and Cram\'{e}r-Rao bounds and discussed the implications of these new limits in the context of quantum metrology and quantum computing. The generalized expressions we obtained do not rely on any specific modification of quantum mechanics and can thus be applied to any system in which such high-order corrections could come into play. As an example, we have considered a specific class of non-local modifications of quantum mechanics inspired by different approaches to quantum gravity. Our analysis not only showed a quantitative reduction of the limits but, above all, revealed a completely new physics (the oscillation in time of the aforementioned bounds) which, according to the above models, provides a low-energy signature of quantum-gravity induced non-locality. We thus expect our work to motivate new experimental tests that could either detect deviations of the above limits or otherwise cast severe constraints on the deviation parameter $\eta$, as well as to shed new light in the physics of some condensed-matter systems, the fluctuations of which can be described in terms of modified effective field theories. The latter, in turn, can be exploited as analog simulators of semi-classical gravity and quantum-gravity phenomenology.

\section{Acknowledgements}
Salman Sajad Wani acknowledges that part of the work was conducted while working under The Scientific and Technological Research Council of Türkiye (TÜBİTAK) BİDEB 2232-A program under project number 121C067. 

\section{Appendix}

In this section we briefly review the theoretical framework at the basis of the example we discussed in the main text. An exhaustive discussion of this topic can be found in Ref. \cite{belen17}.
We consider a free complex massive scalar field, $\phi$, of mass $m$ the dynamics of which is described by the modified Klein-Gordon equation $f(\Box+\mu^2)\phi=0$. Here $f$ is some non-polynomial function of the Klein-Gordon (KG) operator $\Box = c^{-2}\partial^2_t -\nabla^2$, such that $f(\Box+\mu^2)\rightarrow \Box+\mu^2$ in the limit $l_k\rightarrow 0$, where $l_{k}$ is the the characteristic scale of non-locality; $\mu=mc/\hbar$ is the inverse of the reduced Compton wavelength of the field.

Similar nonlocal evolution equations are found in different quantum-gravity models. For example, in four dimensions string field theory predicts a nonlocal KG equation of the form~\cite{sft}
\begin{equation}
f(\Box+\mu^2)=(\Box+\mu^2)\,\exp{[l_k^2(\Box+\mu^2)]}\, .
\label{KGSFT}
\end{equation} 
While $l_k$ is usually identified with the Planck length, in the context of phenomenological models should be safely considered as a free parameter to be bound by the experiments.

We assume that $f$ is an analytic function, so that it can be formally expanded as a power series (this is not generally the case) $f(z) = \sum_{n=1}^\infty b_n z^n$. Upon making the ansatz  $\phi(x)=e^{-i \frac{m c^2}{\hbar}t}\psi(t,x)$ and taking the non relativistic limit ($c\rightarrow\infty$), the nonlocal Klein-Gordon equation reduces to a nonlocal Scr\''odinger equation of the form
\begin{equation}
\label{nlse}
f(\mathcal{S})\psi(t,x) = 0,
\end{equation}
where 
\begin{equation}
\mathcal{S}=i\hbar \frac{\partial}{\partial t} + \frac{\hbar^2}{2m}\nabla^2,
\end{equation}
is the usual Schr\"odinger operator and $f(\mathcal{S})$ can be expanded in power series 
\begin{equation}
f(\mathcal{S})= \mathcal{S}+\sum_{n=2}^{\infty}b_{n}\left(\frac{-2m}{\hbar^{2}}\right)^{n-1}l_k^{2n-2}\mathcal{S}^{n},
\label{scr}
\end{equation}
where the $b_{n}$ are dimensionless coefficients.  

We now focus on the nonlocal dynamics of quantum harmonic oscillator introducing the harmonic potential  $V(x)= \frac{1}{2}m\omega^2 x^2$, $m$ is the mass of the system and $\omega$ its natural angular frequency (we assume that low energy effects of nonlocality only enters in modifying the spacetime derivatives, while not affecting the form of the potential).

The introduction of the potential allows one to construct a dimensionless parameter $\eta \equiv m \omega l_k^2/\hbar$ which can be used to define the perturbative expansion. Note that $\sqrt{\eta}$ represents the ratio between $l_k$ and the width of the oscillator's ground-state wavefunction $x_0 = \sqrt{\hbar/m\omega}$. Using the dimensionless coordinates $t^{'}=\omega t$ and $x^{'}=x/x0$ the nonlocal Scr\''{o} equation for the harmonic oscillator rewrites as
\begin{equation}
\left({\mathcal{S}}+\sum_{n=2}^{\infty}b_{n}\epsilon^{n-1}(-2)^{n-1}{\mathcal{S}}^{n}\right){\psi}=\frac{1}{2}{x}^{2}{\psi}.
\label{schpert}
\end{equation}
We assume solutions of the form
\begin{equation}
\psi= \sum_{n=0}^{\infty}\epsilon^{n}\psi_{n},
\label{expansion}
\end{equation}
where $\psi_{0}$ is a solution to the standard Schr\"odinger equation with harmonic potential,
and higher order terms are suppressed by powers $\epsilon$. In doing this we are implicitly assuming that the non-local equation admits $\psi_{0}$ as a good approximate solutions, assumption which is strongly motivated by experimental observations.

At zeroth-order in $\epsilon$ we have the standard Schr\"odinger equation with a harmonic oscillator 
potential for which we choose a solution corresponding to a coherent state
\begin{equation}
\psi_0=\frac{1}{\pi^{1/4} } \ \exp\left[\sqrt{2} \alpha  
e^{-i t} x-\frac{1}{2} \alpha ^2 e^{-2 i t}-\frac{\alpha ^2}{2}-\frac{i t}{2}-\frac{x^2}{2}\right],
\label{coh}
\end{equation}
where $\alpha$ is the coherent amplitude.
To solve the equation to first order we make the following ansatz
\begin{eqnarray}
\psi_{1}(t,x)= \psi_{0}^{\alpha}(t,x)\left[c_{0}(t)+c_{1}(t) x+c_{2}(t)x^2 \right. \nonumber\\
\left.+c_{3}(t)x^3 +c_{4}(t)x^4 \right].
\end{eqnarray}
Substituting this into Eq. [\ref{schpert}] and keeping only terms of order $\eta$ one finds
\begin{align}\label{coeff}
    & c_{0}(t)=\frac{1}{32} a_{2} \,e^{-8 i t} \left(\alpha ^4-8 \alpha ^4 e^{2 i t}+8 \alpha ^4 e^{6 i t}-\alpha ^4 e^{8 i t}-6 \alpha ^2 e^{2 i t}+20 \alpha ^2 e^{4 i t}\right.\\ \nonumber
    &\quad\quad\quad\qquad\qquad\quad \left. -14 \alpha ^2 e^{6 i t}+28 \alpha ^2 e^{8 i t}-3 e^{4 i t}-4 e^{6 i t}+7 e^{8 i t}\right),\\
    & c_{1}(t)=-\frac{1}{4 \sqrt{2}} \alpha\,  a_{2}\, e^{-7 i t} \left(\alpha ^2-6 \alpha ^2 e^{2 i t}+3 \alpha ^2 e^{4 i t}+2 \alpha ^2 e^{6 i t}-3 e^{2 i t}+4 e^{4 i t}-e^{6 i t}\right),\\
    & c_{2}(t)=\frac{1}{8} \,a_{2}\, e^{-6 i t} \left(3 \alpha ^2-12 \alpha ^2 e^{2 i t}+9 \alpha ^2 e^{4 i t}-3 e^{2 i t}-2 e^{4 i t}+5 e^{6 i t}\right),\\
    & c_{3}(t)=-\frac{1}{2 \sqrt{2}}\, \alpha\,  a_{2}\, e^{-5 i t} \left(1-e^{2 i t}\right)^2,\\
    & c_{4}(t)=\frac{1}{8} a_{2}\, e^{-4 i t} \left(1-e^{4 i t}\right).
\end{align}

\subsection{Time-dependent bounds for quantum gravity-induced nonlocal oscillators}

The Hamiltonian formulation of the above non-local harmonic oscillator model has been derived in Ref. \cite{belen19}.
At first order in the perturbation parameter $\eta$ the Hamiltonian is
\begin{equation}
 H_g =   H+ \eta H_1= \frac{1}{2}\left({x}^2+ {p}^2\right) + \eta \left( \frac{{x}^4}{2} + 2 i x p + 1 \right)
\end{equation}
The average energy of a deformed state can be expressed as  
  \begin{eqnarray} 
  \langle H_g \rangle_g &=& \frac{1}{N} \left(
    \langle\psi(x,t)| H|\psi(x,t)\rangle \right) \nonumber \\
    &+& \frac{\eta}{N}\left(Re[\langle\psi(x,t)|  H_1 |\psi(x,t)\rangle] + \langle \psi_1(x,t)| H|\psi(x,t)\rangle+ \langle\psi(x,t)| H|\psi_1(x,t)\rangle\right) \nonumber \\
\end{eqnarray}

Using Eq. (\ref{coh}) and (\ref{coeff}) we find that the average energy of the deformed coherent state can be expressed as  
  \begin{equation} 
  \langle H_g \rangle_g = \frac{1}{N}\left( E_0 + \eta\left[ E_1 + E_2\right]\right),
\end{equation}
where
\begin{eqnarray}
E_{0} &=& \langle \psi (x,t)| H| \psi (x,t) \rangle = \frac{1}{2} +  \alpha^2
\\
 E_{1}(t) &=&  \frac{1}{16} \left[3 + 6 \alpha^2 (2 + \alpha^2) + 
   2 \alpha^2 ((6 + 4 \alpha^2) \cos(2t) + \alpha^2 \cos(4 t))\right]
\\
E_{2}(t) &=&  \frac{1}{16}  \left[7 + 18 \alpha^2 + 44 \alpha^4 - 2 (1-3 \alpha^2 + 6 \alpha^4) \cos (2t) + (5 + 10 \alpha^2 + 4 \alpha^4) \cos(4t)\right]
\end{eqnarray}
and \begin{eqnarray}
N &=& \langle\psi_g|\psi_g\rangle  
 = 1 +  \frac{\eta}{8} \left( 7 + 12 \alpha^2 + 2 (-1 + \alpha^2 ) \cos(2t) - 5 \cos (4t)\right), \nonumber \\
 &=& 1 + \eta \ n(t)
\end{eqnarray}
is the overall normalization.
%$N= 1 + \eta \ n(t)$ is the overall normalization: 
%\begin{eqnarray}
%N &=& \langle\psi_g|\psi_g\rangle = \int^{\infty}_{-\infty} |\psi_0|^2 (1 + \eta (C_n + C^{*}_n)),  \nonumber \\ 
% &=& 1 +  \frac{\eta}{8} \left( 7 + 12 \alpha^2 + 2 (-1 + \alpha^2 ) \cos(2t) - 5 \cos (4t)\right).
%\end{eqnarray}
Thus we have the total energy
\begin{eqnarray}
    B (t)  &=& \frac{1}{N} \left(E_{0} + \eta [E_{1}(t) + E_{2}(t)]\right) \nonumber \\
    &\approx& E_{0}  + \eta [-E_0 n(t) + E_{1}(t) + E_{2}(t)] \nonumber \\
    &=& E_{0}  + \eta \ B_t ,
    \label{toten}
\end{eqnarray}
where $B_t = - E_{0} \ n(t) + E_1 (t) + E_2(t) $. 

The modified Margolus-Levitin bound can thus be written as 
\begin{equation} \label{bound3}
t_{min}\geq  \frac{\pi}{2}\frac{\hbar}{B (t)} \approx \frac{\pi \hbar}{2 E_0} \left(1-\eta \frac{B_t}{E_0} \right) \ .
\end{equation}
The modified Heisenberg limit is obtained using the above $B_t$ as described in the main text.  

%%%%%%%%%%%%%%%%%%%%%%%%%%%%%%%%%%%%%%%%%%


\begin{thebibliography}{99}
\bibitem{ghirardi86} G.C. Ghirardi, A. Rimini, and T. Weber, Phys. Rev. D {\bf 34}, 470 (1986).
\bibitem{ghirardi90} G.C. Ghirardi, P. Pearle, and A. Rimini, Phys. Rev. A {\bf 42}, 78 (1990).
\bibitem{bassi2003} A. Bassi, and G. C. Ghirardi, Phys. Rep. {\bf 379}, 257 (2003).
\bibitem{bassi2013}A. Bassi, K. Lochan, S. Satin, T. P. Singh, and H. Ulbricht, Rev. Mod. Phys. {\bf 85}, 471 (2013).
\bibitem{penrose96} R. Penrose, Gen. Relativ. Gravit. {\bf 28}, 581 (1996).
\bibitem{diosi} L. Diosi, Phys. Lett. A {\bf 105}, 199 (1984).
\bibitem{1a} M. Carlesso, S. Donadi, L. Ferialdi, M. Paternostro, H. Ulbricht and A. Bassi, Nature Phys.  {\bf 18}, 243 (2022)
\bibitem{1b} A. Vinante, R. Mezzena, P. Falferi, M. Carlesso and A. Bassi,  Phys. Rev. Lett. {\bf 119}, 110401 (2017) 
\bibitem{vinante19} A. Vinante, A. Pontin, M. Rashid, M. Toros, P.F. Barker, H. Ulbricht, Phys. Rev. A {\bf 100}, 012119 (2019).
\bibitem{curceanu} C. Curceanu, B.C. Hiesmayr, and K. Piscicchia, J. Adv. Phys. {\bf 4}, 263 (2015).
\bibitem{class}  L. Petruzziello and  F. Illuminati, Nature Commun. {\bf 12}, 4449 (2021).
\bibitem{veneziano} D. Amati, M. Ciafaloni,  , and  G. Veneziano,  Phys. Lett. B {\bf 197}, 81 (1987).
\bibitem{gross} D. J. Gross and   P. F. Mende,  Nucl. Phys. B {\bf 303}, 407 (1988).
\bibitem{sabine} S. Hossenfelder, Living Rev. Rel.  {\bf 16}, 2  (2013). 
\bibitem{garay}  L. G. Garay, Int. J. Mod. Phys. A {\bf 10}, 145 (1995).
\bibitem{scardigli} F. Scardigli, Phys. Lett. B {\bf 452}, 39 (1999).  
\bibitem{ml12} S. Das and E. C. Vagenas,  Phys. Rev. Lett. {\bf 101}, 221301 (2008).  
\bibitem{ml15} A. F. Ali, S. Das and E. C. Vagenas,  Phys. Rev. D {\bf 84}, 044013 (2011)  
\bibitem{space} A.~F.~Ali, S.~Das and E.~C.~Vagenas, Phys. Lett. B {\bf 678}, 497 (2009). 
\bibitem{space1} S.~Das, E.~C.~Vagenas and A.~F.~Ali, Phys. Lett. B  {\bf 690}, 407  (2010).
\bibitem{ab12} F.~Marin \textit{et al.} Nature Phys. {\bf 9}, 71 (2013); F.~Marin \textit{et al.}, New J. of Phys. {\bf 16}, 085012 (2014). 
\bibitem{ml14}  I. Pikovski, M. R. Vanner, M. Aspelmeyer, M. Kim and C. Brukner,  Nature Phys. {\bf 8}, 393 (2012). 
\bibitem{omt} M. Bawaj \textit{et al.}, Nature Comm. {\bf 6}, 7503 (2015); M. Bonaldi {\it et al.} Eur. Phys. J. D {\bf 74}, 178 (2020).
\bibitem{effect1}M.~Faizal, A.~F.~Ali and A.~Nassar, Phys. Lett. B  {\bf 765}, 238  (2017).
\bibitem{effect2}M.~Faizal, A.~F.~Ali and A.~Nassar, Int. J. Mod. Phys. A  {\bf 30},  1550183 (2015).  
\bibitem{gupa}N.~A.~Shah, A.~Contreras-Astorga, F.~Fillion-Gourdeau, M.~A.~H.~Ahsan, S.~MacLean, and M.~Faizal, Phys. Rev. B {\bf 105}, L161401 (2022).  
\bibitem{gupb}A.~Iorio, P.~Pais, I.~A.~Elmashad, A.~F.~Ali, M.~Faizal and L.~I.~Abou-Salem, Int. J. Mod. Phys. D  {\bf 27}, 1850080 (2018).
\bibitem{girelli08} F. Girelli, S. Liberati and L. Sindoni, Phys. Rev. D {\bf 78}, 084013 (2008).
\bibitem{marino19} F. Marino Phys. Rev. A {\bf 100}, 063825 (2019).
\bibitem{kempf} A. Kempf,  G. Mangano, and R. B. Mann, Phys. Rev. D {\bf 52}, 1108 (1995).
\bibitem{chang} L. N. Chang,  D. Minic, N. Okamura,  and T. Takeuchi, Phys. Rev. D {\bf 65} 125027 (2002).
\bibitem{lewis} Z. Lewis,  and T. Takeuchi,  Phys. Rev. D {\bf 84}, 105029 (2011).
\bibitem{ching} C. L. Ching,  W. K. Ng, Phys. Rev. D {\bf 88} 084009 (2013).
\bibitem{pedram} P. Pedram, Int. J. Mod. Phys. D {\bf 22} 1350004 (2013).
\bibitem{non1}M. Chaichian, M. M. Sheikh-Jabbari, and A. Tureanu, Phys. Rev. Lett. {\bf 86} 2716 (2001).
\bibitem{non2} S. Gangopadhyay and F. G. Scholtz, Phys. Rev. Lett. {\bf 102}, 241602 (2009).
\bibitem{qsl} S. Deffner and  S. Campbell, J. Phys. A: Math. Theor. {\bf 50}, 453001 (2017).
\bibitem{mt}  L. I. Mandelshtam and I. E. Tamm,  J. Phys. {\bf 9},  249 (1945).
\bibitem{ml} N. Margolus and L. B. Levitin, Phys. D {\bf 120}, 188 (1998).
\bibitem{compution} S. Lloyd,  Nature,  {\bf 406}, 1047 (2000).
\bibitem{deffner} S. Deffner, E. Lutz, Phys. Rev. Lett. {\bf 111}, 010402 (2013).
\bibitem{cimmarusti} A. D. Cimmarusti, Z. Yan, B. D. Patterson, L.P. Corcos, L. A. Orozco, S. Deffner, Phys. Rev. Lett. {\bf 114}, 233602 (2015).


\bibitem{orth12}T. Fogarty, S. Deffner, T. Busch and  S.  Campbell, Phys. Rev. Lett. {\bf 124}, 110601 (2020).
\bibitem{symme}I. Marvian, R. W. Spekkens and P. Zanardi, Phys. Rev. A {\bf 93}, 052331 (2016).
\bibitem{kz12}R. Puebla, S. Deffner and S, Campbell, Phys. Rev. Research {\bf 2}, 032020(R)   (2020).
\bibitem{2001a} Y. Jack Ng, Phys. Rev. Lett. {\bf 86}, 2946  (2001).
\bibitem{foam1}S. Carlip, Phys. Rev. Lett. {\bf 79}, 4071 (1997).
\bibitem{foam5} A. Perez and D. Sudarsky, Phys. Rev. Lett. {\bf 122}, 221302  (2019). 
\bibitem{foam2} W. A. Christiansen, Y. Jack Ng, and H. van Dam, Phys. Rev. Lett. {\bf 96}, 051301 (2006). 
\bibitem{foam4} W. A. Christiansen, Y. J. Ng, David J. E. Floyd and E. S. Perlman, Phys. Rev. D {\bf 83}, 084003 (2011).  
\bibitem{1995}L. H. Ford, Phys. Rev. D {\bf51}, 1692 (1995). 
\bibitem{ligo1} Y.~J.~Ng and H.~van Dam, Phys. Lett. B {\bf 477}, 429-435 (2000).
\bibitem{ligo2} G.~Calcagni, S.~Kuroyanagi, S.~Marsat, M.~Sakellariadou, N.~Tamanini and G.~Tasinato,
Phys. Lett. B {\bf 798}, 135000 (2019). 

\bibitem{miro1}M.~Faizal, M.~M.~Khalil and S.~Das, Eur. Phys. J. C {\bf 76}, 30 (2016). 
\bibitem{miro9}B.~G.~Sidharth, Eur. Phys. J. C {\bf 76}, 206 (2016).
\bibitem{miro2} S.~Aghababaei, H.~Moradpour, G.~Rezaei and S.~Khorshidian, Phys. Scripta {\bf96}, 055303 (2021).
\bibitem{miro4} F.~Feleppa, H.~Moradpour, C.~Corda and S.~Aghababaei, EPL {\bf 135}, 40003 (2021).
\bibitem{miro6} S. A. Alavi, Phys. Scripta  {\bf 67}, 366-368 (2003).  

\bibitem{quant1} M. Zwierz, C. A. Perez-Delgado, and P. Kok, Phys. Rev. Lett. {\bf 105}, 180402 (2010).
\bibitem{est1} S. L. Braunstein and C. M. Caves, Phys. Rev. Lett. {\bf 72}, 3439 (1994).
\bibitem{est2} S. L. Braunstein, C.  M. Caves and G. J. Milburn, Annals  Phys. {\bf 247}, 135 (1996).
\bibitem{nu} V. Giovannetti, S. Lloyd and L. Maccone, Phys. Rev. Lett. {\bf 108}, 260405 (2012).

\bibitem{qc1} W.~Cottrell and M.~Montero, JHEP  {\bf 02}, 039 (2018).
\bibitem{qc2} W.~C.~Gan and F.~W.~Shu, Phys. Rev. D  {\bf 96},  026008 (2017).
\bibitem{rev} C. Barcel\`o, S. Liberati, M. Visser, Living Rev. Rel. {\bf 14}, 3 (2011).
\bibitem{carusotto2008} I. Carusotto, S. Fagnocchi, A. Recati, R Balbinot, and A. Fabbri, New J. of Phys. {\bf 10} 103001 (2008).
\bibitem{steinhauer2014} J. Steinhauer, Nat. Phys. {\bf 10}, 864 (2014).
\bibitem{steinhauer2016} J. Steinhauer, Nat. Phys. {\bf 12}, 959 (2016).
\bibitem{weinfurtner2011} S. Weinfurtner, E. W. Tedford, M. C. J. Penrice, W. G. Unruh, G. A. Lawrence, Phys. Rev. Lett. {\bf 106}, 021302 (2011).
\bibitem{drori} J. Drori, Y. Rosenberg, D. Bermudez, Y. Silberberg, and U. Leonhardt Phys. Rev. Lett. {\bf 122}, 010404 (2019).
\bibitem{prain2019} A. Prain, C. Maitland, D. Faccio, and F. Marino, Phys. Rev. D {\bf 100}, 024037 (2019).
\bibitem{torres} T. Torres, S. Patrick, A. Coutant, M. Richartz, E. W. Tedford and S. Weinfurtner, Nat. Phys. {\bf 13}, 833 (2017).
\bibitem{braidotti2022} M. C. Braidotti {\it et al.} Phys. Rev. Lett. {\bf 128}, 013901 (2022).
\bibitem{ciszak} M. Ciszak and F. Marino Phys. Rev. D {\bf 103}, 045004 (2021).
\bibitem{hod} S. Hod, Phys. Rev. D {\bf 104}, 104041 (2021).
\bibitem{belen16} A.~Belenchia, D.~M.~T.~Benincasa, S.~Liberati, F.~Marin, F.~Marino and A.~Ortolan,
Phys. Rev. Lett.  {\bf 116}, 161303 (2016).
 \bibitem{belen17} A.~Belenchia, D.~M.~T.~Benincasa, S.~Liberati, F.~Marin, F.~Marino and A.~Ortolan,
Phys. Rev. D  {\bf 95}, 026012 (2017).
\bibitem{eliezer} D. A. Eliezer and R. P. Woodard, Nucl. Phys. B {\bf 325}, 389 (1989).
\bibitem{szabo} R. J. Szabo, Phys. Rep. {\bf 378}, 207 (2003).
\bibitem{sorkin} R. D. Sorkin, in Approaches to Quantum Gravity: Towards a New Understanding of Space and Time, edited by D. Oriti (Cambridge University Press, Cambridge, England, 2006).
\bibitem{gambini}R. Gambini and J. Pullin, Int. J. Mod. Phys. D {\bf 23}, 1442023 (2014).
%\bibitem{bonaldi} M. Bonaldi {\it et al.} The EPJ D {\bf 74}, 178 (2020).
\bibitem{belen19} A.~Belenchia, D.~M.~T.~Benincasa, F.~Marin, F.~Marino, A.~Ortolan, M.~Paternostro and S.~Liberati, Class. Quant. Grav. {\bf 36}, 155006 (2019).
 
\bibitem{peterson} R. W. Peterson, T. P. Purdy, N. S. Kampel, R. W. Andrews,
P. L. Yu, K. W. Lehnert, and C. A. Regal, Phys. Rev. Lett. {\bf 116}, 063601 (2016).
\bibitem{Qiu} L. Qiu, I. Shomroni, P. Seidler, and T. Kippenberg, Phys. Rev. Lett. {\bf 124}, 173601 (2020).
\bibitem{Jones2010} P. J. Jones, and P. Kok, Phys. Rev. A {\bf 82}, 022107 (2010).
\bibitem{m2} M. Zwierz, Phys.  Rev.  A {\bf 86}, 016101 (2012).
\bibitem{note} Even though the original Margolus–Levitin bound was obtained for orthogonal states, the result can be generalized to any state \cite{Jones2010, m2}.  
\bibitem{sft} A. S. Koshelev, Rom. J. Phys. {\bf 57}, 894 (2012).
\end{thebibliography}
\end{document}